\documentclass[12pt]{iopart}
\usepackage{graphicx}
\begin{document}
\title[Cosmic Sculpture]{Cosmic Sculpture: A new way to visualise the Cosmic Microwave Background}
\author{D. L. Clements$^1$, S. Sato$^{1}$,
A. Portela Fonseca$^1$}
\address{$^1$ Department of Physics, Blackett Lab.,
Imperial College, Prince Consort Road, London SW7~2AZ, UK}
\ead{d.clements@imperial.ac.uk}
\begin{abstract}
3D printing presents an attractive alternative to visual representation of physical datasets such as astronomical images that can be used for research, outreach or teaching purposes, and is especially relevant to people with a visual disability. We here report the use of 3D printing technology to produce a representation of the all-sky Cosmic Microwave Background (CMB) intensity anisotropy maps produced by the {\em Planck} mission. The success of this work in representing key features of the CMB is discussed as is the potential of this approach for representing other astrophysical data sets. 3D printing such datasets represents a highly complementary approach to the usual 2D projections used in teaching and outreach work, and can also form the basis of undergraduate projects. The CAD files used to produce the models discussed in this paper are made available. 
\end{abstract}

\submitto{\EJP}
\maketitle

\section{Introduction}

Observations of anisotropies in the Cosmic Microwave Background (CMB) have been fundamental to the development of our understanding of the formation and development of the Universe. Starting with the COBE satellite (Boggess et al., 1992) which provided the first detection of the intrinsic CMB temperature fluctuations, there have been a series of ground and space-based experiments that have measured these fluctuations to ever increasing precision and on an ever-widening range of angular scales (eg. Jaffe et al., 2001; Bennett et al., 2003). The most recent CMB experiment was the {\em Planck} mission, which has produced a map of the CMB temperatures fluctuations over the entire sky down to angular scales of $\sim$ 7 arcminutes (Fig. 1, Planck Collaboration, 2014a).

\begin{figure}
\includegraphics[width=15cm]{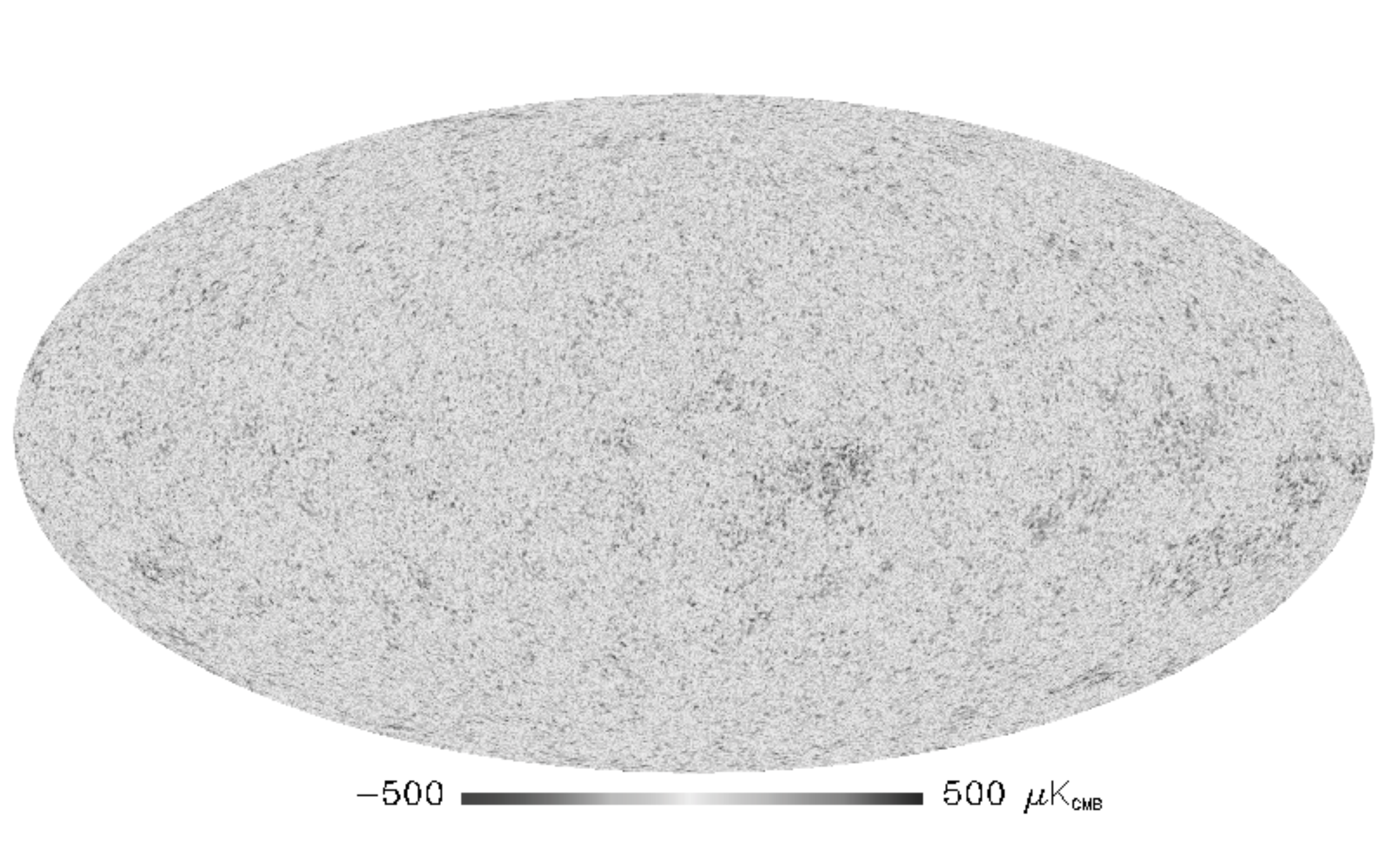}
\caption{The all sky CMB anisotropy map from {\em Planck} (Planck Collaboration, 2014a, ESA).}
\label{fig:cmb}
\end{figure}

CMB temperature fluctuations are the result of acoustic oscillations in the plasma of the early universe that were frozen into the background radiation when matter and radiation decoupled at the time of recombination. The statistics of the fluctuations are Gaussian with a typical relative amplitude of $\Delta T/T \sim 10^{-5}$ superimposed on the $T = 2.73$ K black body radiation.
The strength of fluctuations at different angular scales is given by their power spectrum (Fig. 2, Planck Collaboration, 2014a) which shows a clear peak on scales of $\sim$ 1 degree, which corresponds to the horizon scale at the time of recombination. The shape of the CMB power spectrum is a key tool in constraining cosmological models.
\begin{figure}
\includegraphics[width=15cm]{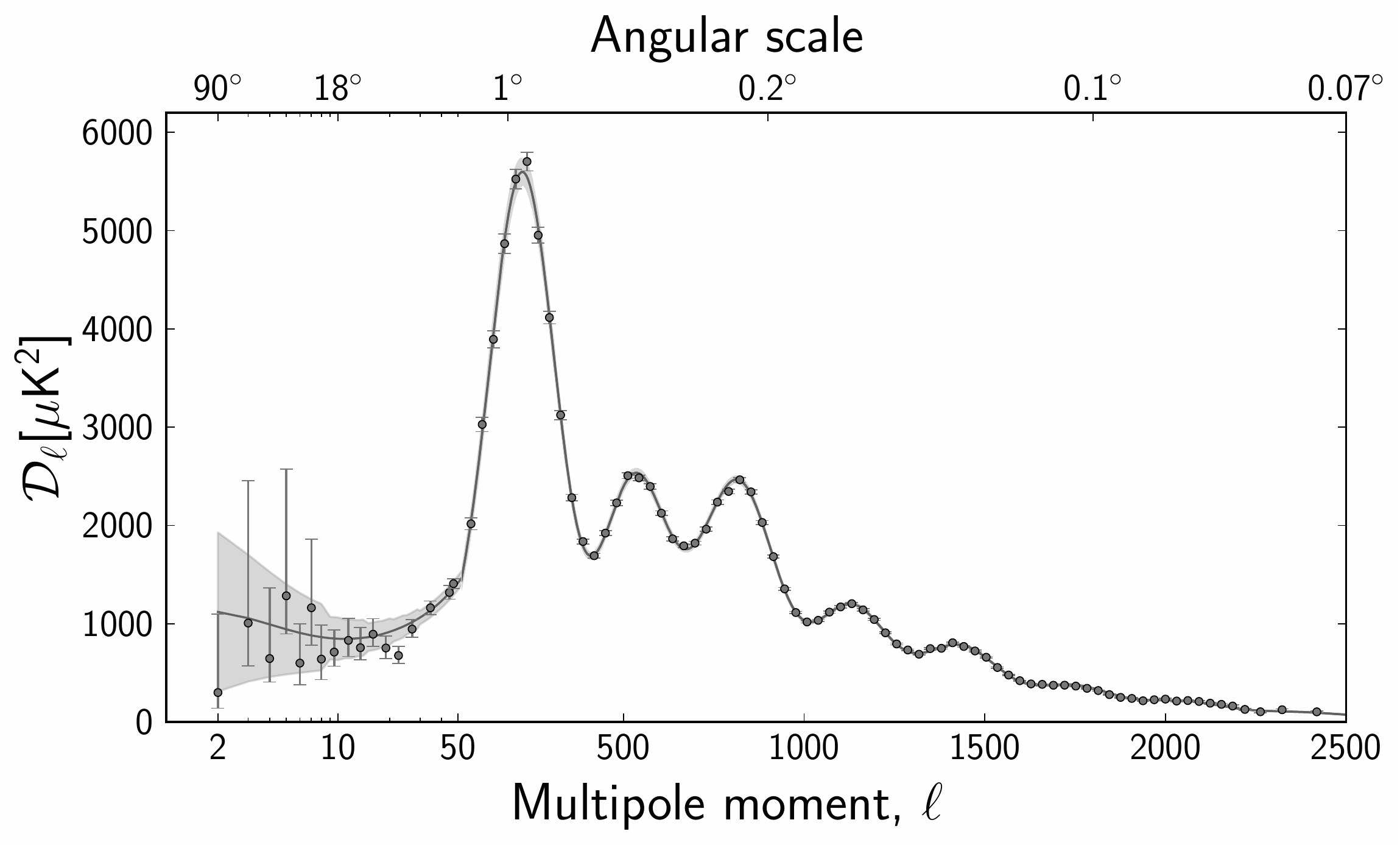}
\caption{The CMB power spectrum derived from the {\em Planck} CMB map (Planck Collaboration, 2014a, ESA).}
\end{figure}

The traditional way that CMB fluctuations are visualised is to represent the CMB temperature range with a colour table - eg. red for the hottest scaling to blue for the coldest - and then to display these colours on a sheet of paper using one of the standard spherical projections, most often a Mollweide projection (see Fig. 1). Some attempts have been made to produce interactive computer graphics visualisations of the CMB as a sphere, which can then be rotated at random by a user, but all of the representations to date have relied upon vision as the key tool for representing the varying temperature of the CMB. In this paper we describe how an alternative approach to representing the CMB temperature fluctuations can be achieved by using a 3D printer to render the anisotropies as bumps and dips on the surface of a sphere. Presenting the CMB in a truly 3D form, that can be held in the hand and felt rather than viewed, has many potential benefits for teaching and outreach work, and is especially relevant for those with a visual disability. While 3D printers have previously been used to visualise mathematical functions (Aboufadel et al., 2013; Knill \& Slavkovsky, 2013) and the results of models of complex systems (Reiss et al., 2013), its use to represent astronomical observational data is new, and is potentially a novel way for researchers to represent and interrogate their data as well as to present it to students. 3D printing is also of use for outreach in that it allows those with impaired vision to better understand and appreciate the results of observational cosmology. It also permits everyone to use a different sense, that of touch, to appreciate the properties of the CMB.

This paper describes the use of 3D printers to represent cosmological observations in this way, but a similar approach can be applied to other astrophysical data sets, including, but not limited to, topographic maps of planets, observations and models of the surface and interior structures of stars, the distribution of stars in galaxies, and the distribution of galaxies and intergalactic gas within the large scale structure of the universe. This work can be regarded as a proof-of-concept demonstration for all of these other possibilities. Since it was the result of a final year undergraduate project at Imperial College London (S. Sato \& and A. Portela Fonseca were the students), this work also serves as a demonstration that this kind of project can play a strong pedagogical role, introducing students to real astronomical or cosmological datasets and to the techniques of 3D printing.

The rest of this paper is structured as follows. In Section 2 we describe the Planck dataset that was used as the starting point for our 3D printing. In Section 3 we discuss the capabilities and limitations of different 3D printing technologies for this work, while in Section 4 we describe how the Planck data were processed to make it appropriate for 3D printing. In Section 5 we describe and present the final products from this work. We draw our conclusions in Section 6.

\section{The {\em Planck} Data}

The {\em Planck} mission observed the entire sky at nine different frequencies, from 30GHz to 857GHz (Planck Collaboration, 2014a). This allowed various foregrounds from our own and other galaxies, including dust, synchrotron radiation and emission lines from CO, that would contaminate the 2.73K CMB emission, to be determined and removed (Planck Collaboration, 2014b). The CMB dipole anisotropy (Smoot et al., 1977), that results from our motion relative to the CMB frame, was also removed, allowing the primary CMB anisotropies to be determined. These anisotropies are of order $\Delta T/T \sim 10^{-5}$ in size.

The all-sky {\em Planck CMB} maps, as well as the individual frequency maps, component separation maps and more, are available online at \verb$http://pla.esac.esa.int/pla/$, where the all-sky maps are stored in HEALPix format\footnote{http://healpix.sourceforge.net/} (Gorski et al., 2005). This is a hierarchical equal area pixellation scheme for a sphere and is the the main data format used for astrophysical all-sky mapping projects like {\em Planck} and {\em WMAP}. HEALPix divides the sphere up into equal area pixels starting with twelve equal area curvilinear quadrilaterals and then, for higher resolutions, dividing these quadrilaterals into ever smaller sets of four smaller quadrilaterals (see Fig. 3 of Gorski et al. (2005)). A given HEALPix resolution is defined by the number of pixels each of the original quadrilaterals has on a side, going up by a factor of 2 at each resolution increase. The native HEALPix scale for the {\em Planck} maps is Nside=2048, corresponding to an angular resolution of each HEALPix pixel of $\sim$ 1.7 arcminutes, and a total number of pixels over the whole sky of 50,331,648.

\section{3D Printing}

3D printing is the term usually applied to a wide range of additive manufacturing techniques whereby a solid object is produced through the computer-controlled addition of material that is typically built up layer by layer. A wide range of materials can be used for 3D printing, ranging from sugar to titanium.

Two materials, and two 3D printing technologies, were used for this project. The first used polylactic acid (PLA), a biodegradable thermosoftening plastic. This is printed layer on layer by the {\em Ultimaker} 3D printer used here. PLA is a cheap printing material and as such was ideal for prototyping and testing the computer aided design (CAD) files developed for this project. The second material, used for the final coloured objects, was plaster, printed by a ZPrinter 650 machine. This printer works in a somewhat different way to the PLA printing machine. While the {\em Ultimaker} only lays down material that will be part of the final object, the ZPrinter operates by printing a binding fluid, which may include colouration, onto a preexisting layer of powder. The entire printing bed of the machine is covered in powder, one layer at a time, but only the material bound together by the printing head is retained in the final product. This means that the object being printed, no matter its shape, is supported throughout the printing process by the powder layers. In contrast, material printed by the {\em Ultimaker} must be self supporting. Since we aim to print a representation of the all-sky CMB as bumps on a sphere, a support structure must be added to the {\em Ultimaker} print to ensure that the bottom part of the sphere, where there is a large overhang, is not in danger of slumping. 

Each printer requires an input file in its own specific format. For the {\em Ultimaker} a stereolithography (STL) file is the starting point. This is then processed through a slicing programme called \verb|Cura| which produces the GCode file used by the printer. The ZPrinter in contrast uses a Virtual Reality Modelling Language (VRML) file as input instead. This provides an alternative way of describing the object to be printed and adds the possibility of colouring the final product. In principle any colour scale can be produced, but for this project a 23 colour scale based on that used for all-sky CMB images in publications by the {\em Planck} consortium was used.

\section{From {\em Planck} Data to CAD Files}

A number of processing steps are needed to go from the initial nside=2048 HEALPix maps to the CAD files needed by each of the 3D printer types used. The first step involves using the HEALPix data to produce a representation of the CMB anisotropies as bumps on the surface of a sphere with the temperature anisotropies scaled by an appropriate amount so that they are large enough to be felt and seen. This spherical representation then has to be processed to produce appropriate input files for the relevant printers.

\subsection{HEALPix to a Sphere}

The original {\em Planck} maps are stored as HEALPix Nside = 2048 objects, corresponding to a resolution of $\sim$1.7 arcminutes and 50,331,648 pixels. The resulting 3D printer file that would be produced from this would be unmanageably large, and would include detail at a resolution finer than can easily be perceived even for a large diameter finally printed sphere - for a 10 cm diameter printed sphere the smallest scale features would be $\sim$50 $\mu$m across, roughly equivalent to the width of a human hair. Furthermore, most of the power in the CMB anisotropies is on scales of $\sim$ 1 degree (see Fig. 1). The CMB HEALPix maps were therefore degraded by a factor of sixteen to HEALPix Nside = 128 maps, corresponding to a pixel size of $\sim$ 0.46 degrees which is thus well matched to the scale at the peak of the CMB anisotropy power spectrum. The total number of HEALPix pixels at this resolution is a much more manageable 196608.

The input CAD files for the 3D printers used in this project consist of a mesh of triangles that make up the surface of the object being printed. Such a mesh can be generated from a list of Cartesian coordinates that lie on the surface of the object by software such as the \verb!MeshLab! system\footnote{http://meshlab.sourceforge.net/}. The next step in producing our 3D representation of the CMB is thus to use the Nside = 128 HEALPix map to produce a list of Cartesian coordinates lying on the surface of the object to be printed.
To do this, the $r, \theta, \phi$ polar coordinates of each HEALPix pixel were calculated. The $r$ coordinate was then modified on the basis of the CMB temperature anisotropy, scaled by a variable factor so that the structures on the surface representing them would be large enough to be discernible by touch or vision. The initial test printings were done using a 3 cm radius sphere, so the radius $r$ was scaled as follows:
\begin{equation}
r(\theta, \phi) = 3 + \left( n \times 1000 \times \Delta T(\theta, \phi) \right)
\end{equation}
where the eventual radial coordinate is in cm, $\Delta T(\theta, \phi)$ is the CMB anisotropy at the angular position ($\theta$, $\phi$) given by the HEALPix map, and $n$ is a scale factor that was tuned to produce the best effect. Values of 1, 2 and 3 were tested for $n$, with 2 being found to produce the most effective results. The list of $(r, \theta, \phi)$ coordinates generated in this way for the surface of the sphere were then converted to $(x,y,z)$ Cartesian coordinates to provide the input to the next stages of the process which convert this list of points into the meshes appropriate for each type of 3D printer.

\subsection{Making the Mesh}

The mesh, linking the points in the $(x,y,z)$ coordinate list and which is later used to produce the CAD files that will be printed, was generated using the \verb!MeshLab!\footnote{http://meshlab.sourceforge.net/} package. This takes the list of points, finds neighbouring points, and then generates a mesh made up of triangles that connect these points using a Poisson surface reconstruction algorithm (Kazhdan et al., 2006). The end result is the surface of the object that will be printed.

To produce the mesh from the input $(x,y,z)$ file,~\verb!MeshLab! uses a tree searching algorithm where the tree depth specifies the the level of detail in the final product. A variety of depths was tried with fourteen finally being selected for the $n=2$ model that was eventually used for printing. The object descriptions produced by  \verb!MeshLab! are not guaranteed to be free of holes or degeneracies. Such flaws were identified and corrected using the \verb!NetFabb!\footnote{http://www.netfabb.com/basic.php} package. The final resulting object description was output to both STL and VRML files, which are appropriate for the {\em Ultimaker} and ZPrinter devices respectively.

\subsection{STL File Generation for the {\em Ultimaker}}

\begin{figure}
\includegraphics[width=15cm]{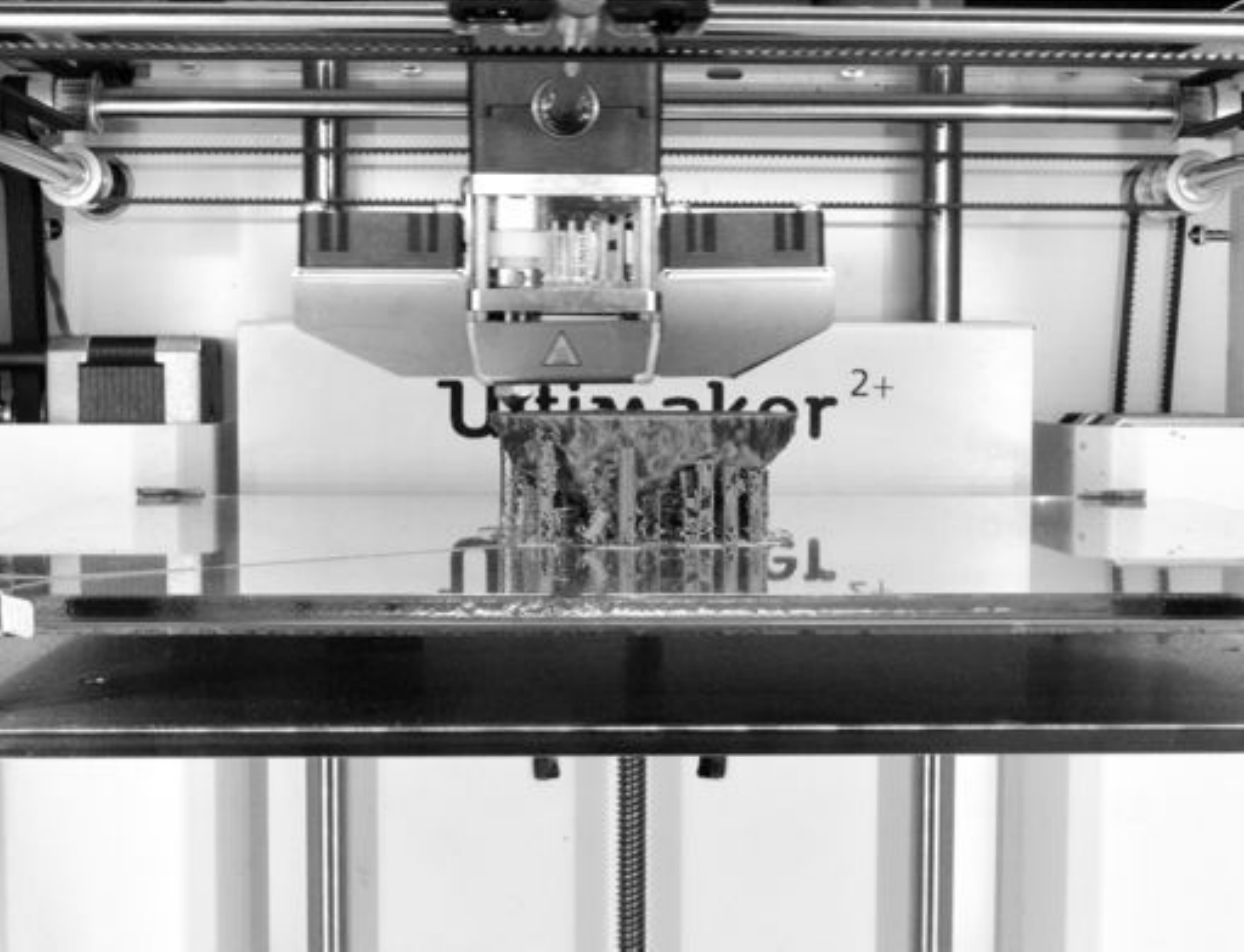}
\caption{3D CMB model being printed by an {\em Ultimaker}. The brown material is the printed plastic, with the print head working above it. The spherical CMB model is supported by a number of vertical supports which can be removed after printing.}
\label{fig:ultimaker}
\end{figure}

The {\em Ultimaker} printer uses the plastic PLA as its printing material, and prints by extruding this plastic onto the printing surface. An overhanging structure, such as the base of a sphere, will  not be able to support itself when printed in this manner. The software \verb!Cura!\footnote{https://ultimaker.com/en/products/cura-software}, is used to further process the input STL files to slice the object into the thin layers that the machine will print. \verb!Cura! can also be used to generate support struts to prevent drooping. \verb!Cura! was also used to give the object a 20\% filled internal grid making the final plastic printed models less breakable. The eventual result of this process is a GCode file which can be input to the {\em Ultimaker}. Figure \ref{fig:ultimaker} shows an {\em Ultimaker} printer in the process of printing our 3D representation of the CMB, complete with \verb!Cura!-generated support structures and a base.

\subsection{VRML File Generation for the ZPrinter}

The printing method for the ZPrinter, where layers of plaster are put down and the print head sprays a fluid that glues the powder together that will form the final object, has several advantages for the current work. Firstly the powder layers provide support during the printing process so no additional support struts are required. Secondly, ink-jet printer ink can be mixed with the fixing fluid so that the final object can be coloured. There are several disadvantages though. Firstly the feedstock for the ZPrinter is more expensive, so the final model is over 10 times the price of the plastic Ultimaker version, and secondly the powder is quite dense so the final model needs to be hollow and have a hole in the surface to allow the interior powder to escape. For these reasons the ZPrinter was only used for our final printed model, with a somewhat larger size than the original 3 cm radius plastic models.

The \verb!MeshLab!-produced mesh was rescaled by a factor of 5/3 using \verb!NetFabb! so that the final model will have a 5 cm radius and then hollowed out to have a shell that is at least 5mm thick. The \verb!Blender!\footnote{https://www.blender.org/} tool was then used to produce a cylindrical hole of diameter 10mm at the south galactic pole of the model. The final mesh was exported from \verb!NetFabb! as VRML, the file format used by the ZPrinter. Before this was done a colour value was defined for each vertex using a 23 colour scale modelled on the \verb!Commander! colour scheme used by the {\em Planck} consortium, where blue represents the coldest points and red the hottest. Figure \ref{fig:vrml} shows the full colour CMB model printed by the ZPrinter.

\begin{figure}
\includegraphics[width=15cm]{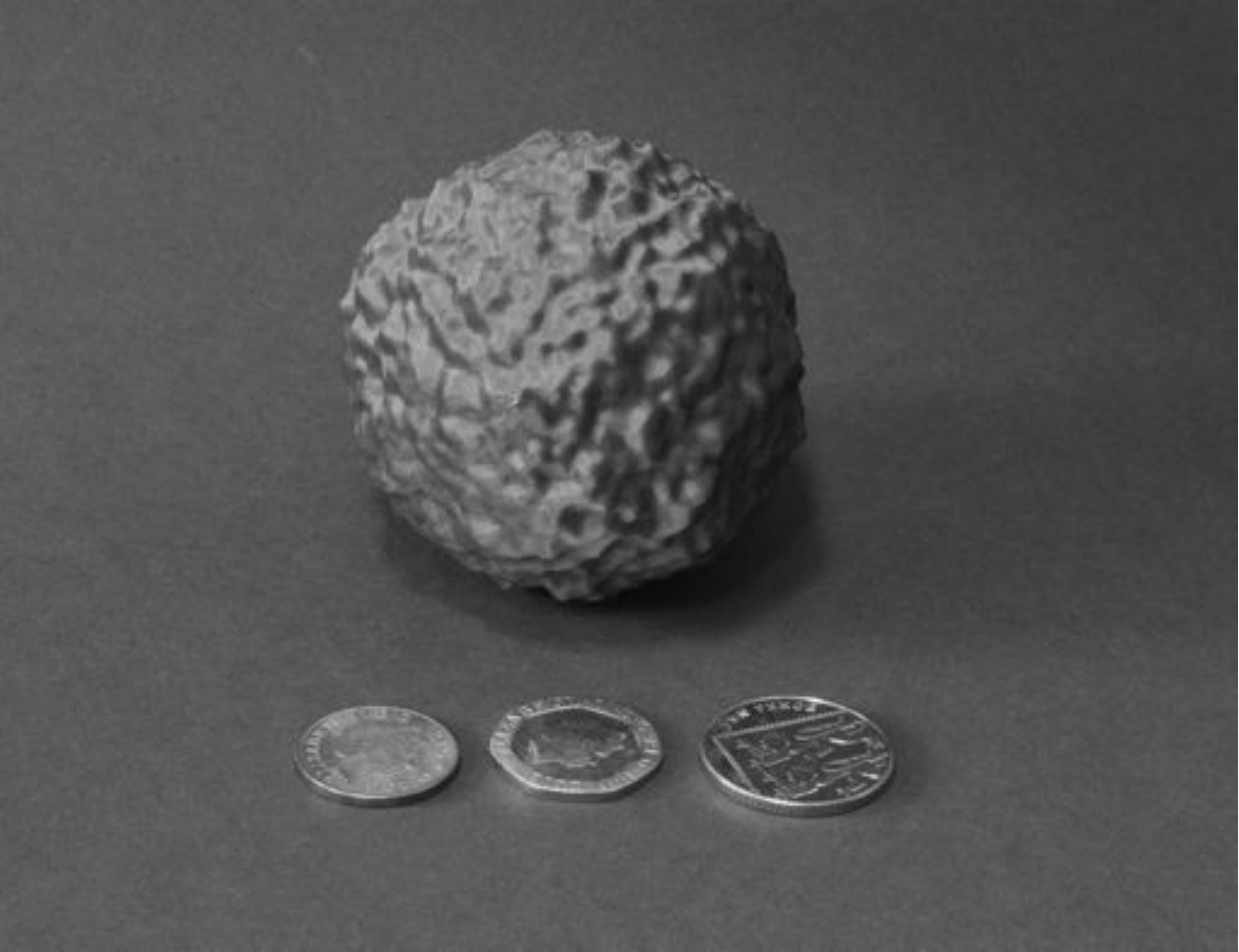}
\caption{The final {\em Ultimaker} printed CMB model.}
\label{fig:ultimaker_cmb}
\end{figure}

\begin{figure}
\includegraphics[width=15cm]{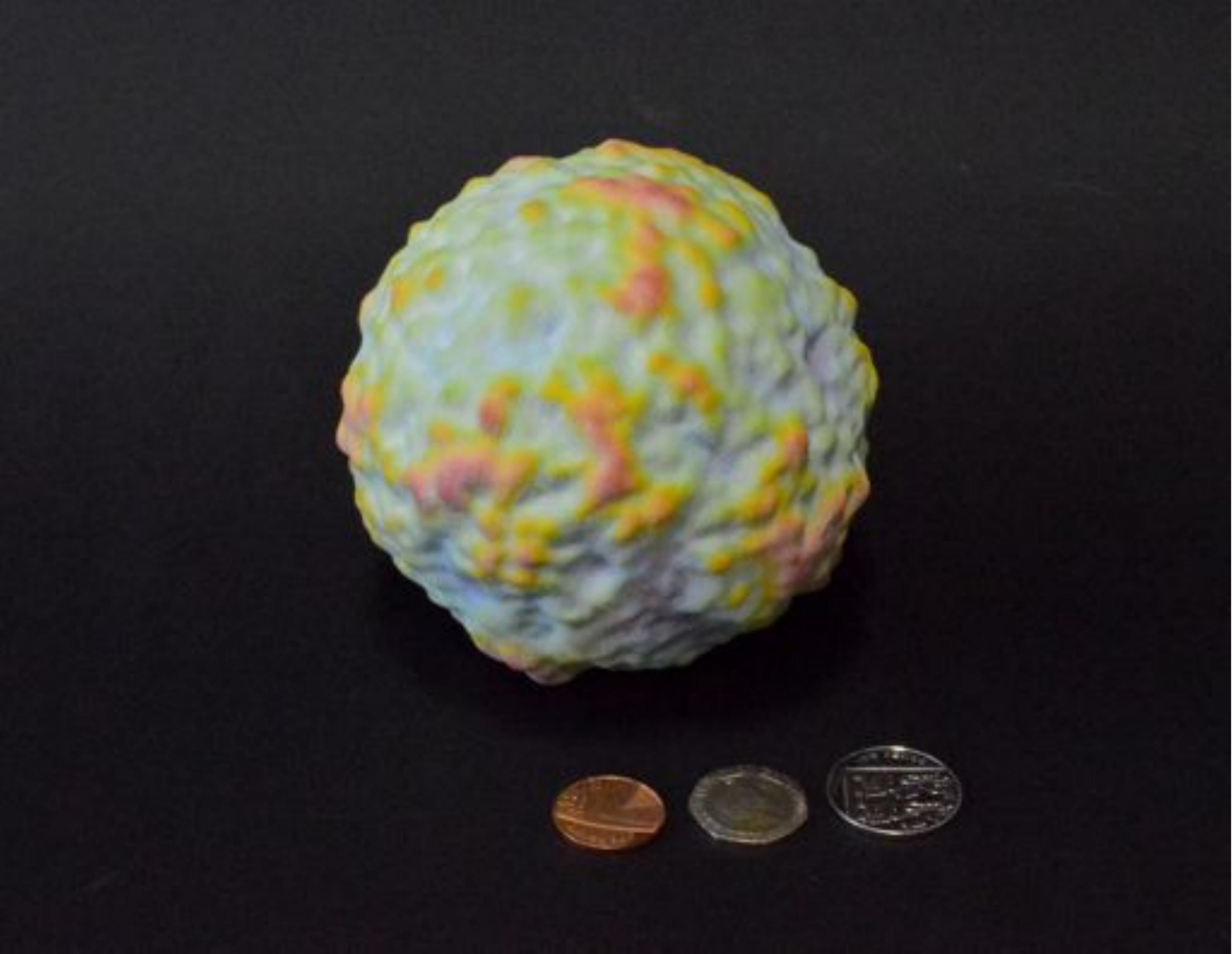}
\caption{The final ZPrinter printed CMB model with colour.}
\label{fig:vrml}
\end{figure}

\section{Printed Universes}

Figures \ref{fig:ultimaker_cmb} and \ref{fig:vrml} show the final printed versions of the CMB models from the ZPrinter and {\em Ultimaker} respectively. As can be seen the structure of the CMB fluctuations on scales matching the first doppler peak are rendered as visible bumps and dips in the surface of the resulting sphere, while the colours of the ZPrinter version are similar to those of the {\em Planck} CMB map in Figure \ref{fig:cmb}. What is less easy to discern from an image is that these structures also produce a highly tactile surface, the features of which are readily appreciated by touch alone.  The highest peaks and, even more easily, the deepest holes, are easily found. The famous `CMB cold spot' (Cruz et al., 2007), an unusually low temperature region in the CMB, first detected by {\em WMAP} and since confirmed by {\em Planck}, can be felt as a small but isolated depression.

These models, though, are not perfect, but they do represent a successful proof of concept for 3D printed representations of astronomical or, in this case, cosmological systems. Since we have restricted ourselves to an Nside=128 HEALPix resolution, much of the improved resolution from the {\em Planck} mission is lost. Nevertheless we retain the bulk of the CMB anisotropy power spectrum around the one degree scale. Detail is also lost through a number of approximations made during the construction of the mesh. These were necessitated by power and memory constraints on the machines used in the mesh construction. Alternative algorithms or a deeper tree search in the selection of neighbouring points, for example, might lead to improvements in the future.

\subsection{Availability of CAD Files}

The CAD files used to produce both the plastic (STL) and plaster (VRML) versions of the CMB are made available with this paper. They can be downloaded from \verb!http://dx.doi.org/10.5281/zenodo.60215! and are made available under a Creative Commons Attribution-NonCommercial-ShareAlike 4.0 International License.

\section{Conclusions}

We have produced a solid 3D representation of the CMB temperature anisotropies using the {\em Planck} mission data and 3D printers. The CAD files that describe the printed objects were produced using a variety of tools, including \verb! MeshLab, Cura, Blender! and \verb!Netfabb! which produced a file in STL for printing on an {\em Ultimaker} machine, and in VRML, for printing on a ZPrinter machine. The final 3D printed objects capture the essence of the CMB anisotropies on angular scales of about 1 degree, matching the peak in the CMB anisotropy power spectrum. They can provide a non-visual appreciation of the CMB for both the sighted and unsighted. The CAD files used to produce these models are freely available for non-commercial use from \verb!http://dx.doi.org/10.5281/zenodo.60215!.

\ack
This paper is based on observations obtained with {\em Planck} (http://www.esa.int/{\em Planck}), an ESA science mission with instruments and contributions directly funded by ESA Member States, NASA, and Canada. This work used \verb!MeshLab!, a tool developed with the support of the 3D-CoForm project. The authors would like to thank the Imperial College Advanced Hackspace, the Imperial College Department of Design Engineering, Sam McKenney, Larissa Kunstel-Tabet, Meilin Sancho, Andrew Jaffe and Tim Evans for their help with this work. It was funded in part by STFC Grant Reference ST/K001051/1.

\References

\item Aboufadel, E, Krawczyk, S V, Sherman-Bennett, M 2013, arXiv:1308.3420
\item Bennett, C K, et al. 2003 {\it ApJS} {\bf 148} 1
\item Boggess, N W, et al., 1992 {\it ApJ} {\bf 397} 420
\item Cruz, M, et al., 2007 {\it ApJ} {\bf 655} 11
\item Gorski, K M, et al., 2005 {\it ApJ} {\bf 622} 759
\item Jaffe, A H, et al., 2001 {\it PhRvL} {\bf 86}, 3475
\item Kazhdan M, Bolitho M \& Hoppe H. Poisson surface reconstruction, Proceeding of the fourth Eurographics symposium on Geometry processing, editors Polthier K, Sheffer A, publisher Eurographics Association, 2006; Vol. 7
\item Knill, O, Slavkovsky, E, 2013, arXiv:1306.5599
\item Planck Collaboration, 2014a {\it A\&A} {\bf 571} 1
\item Planck Collaboration, 2014b {\it A\&A} {\bf 571} 12
\item Reiss, D S, Price, J J, Evans, T, 2013, {\it EPL} {\bf 104}, 48001 
\item Smoot, G F, Gorenstein, M V, Muller, R A,, 1977, {\em PhRvL} {\bf 39} 898

\endrefs

\end{document}